\documentclass[runningheads]{llncs}

\usepackage{eccv}

\usepackage{eccvabbrv}

\usepackage[utf8]{inputenc} 
\usepackage[T1]{fontenc}    
\usepackage{graphicx}
\usepackage{url}            
\usepackage{booktabs}       
\usepackage{amsfonts}       
\usepackage{nicefrac}       

\usepackage{cite}
\usepackage{color}
\usepackage{wrapfig}
\usepackage{epsfig}
\usepackage{graphicx}
\usepackage{amsmath}
\usepackage{amssymb}
\usepackage{multirow}
\usepackage{ulem}

\usepackage{arydshln}
\usepackage{threeparttable}
\usepackage{colortbl}
\usepackage{enumitem}
\usepackage{siunitx}
\usepackage{bm}
\usepackage{array}
\usepackage{caption}
\usepackage{dblfloatfix}
\usepackage{pifont}

\usepackage{float}

\usepackage{listings}
\usepackage[export]{adjustbox}
\usepackage{makecell}
\usepackage{centernot}
\usepackage{math_symbols}
\usepackage{bm}
\usepackage{enumitem}
\usepackage[linesnumbered,ruled]{algorithm2e}
\usepackage{authblk}

\usepackage[accsupp]{axessibility}  

\definecolor{mygray}{gray}{.9}
\definecolor{ggray}{RGB}{127,127,127}
\definecolor{myred}{RGB}{192,0,0}
\definecolor{redb}{RGB}{217,148,143}
\definecolor{myyellow}{RGB}{190,144,0}
\definecolor{mygreen}{RGB}{80,100,40}
\definecolor{myblue}{RGB}{30,90,100}

\def\eg{\textit{e.g}\onedot} 
\def\ie{\textit{i.e}\onedot} 
 
\def\etc{\textit{etc}\onedot} 
 
\def\etal{\textit{et al}\onedot}

\makeatletter
\newcommand{\thickhline}{
    \noalign {\ifnum 0=`}\fi \hrule height 1pt
    \futurelet \reserved@a \@xhline
}

\usepackage{hyperref}

\usepackage{orcidlink}

\begin{document}

\title{Mutual Learning for Acoustic Matching and Dereverberation via
Visual Scene-driven Diffusion} 

\titlerunning{MVSD}

\author{Jian Ma\inst{1,3} \and
Wenguan Wang\inst{2}$^\dag$ \and
Yi Yang\inst{2} \and
Feng Zheng\inst{1}
}

\authorrunning{J. Ma~et al.}

\institute{$^1$Southern University of Science and Technology \\
$^2$ReLER, CCAI, Zhejiang University~~$^3$ReLER, University of Technology Sydney \\
\href{https://hechang25.github.io/MVSD}{\textcolor[rgb]{0.80,0.00,0.25}{https://hechang25.github.io/MVSD}}
}
\maketitle
\begingroup\def\thefootnote{\dag}\footnotetext{\small Corresponding author: Wenguan Wang.}\endgroup
\begin{abstract}
$\!$\textit{Visual acoustic matching~(VAM)} is pivotal for enhancing the immersive experience, 
and the task of \textit{dereverberation}$\!$ is effective in improving audio intelligibility.
Existing methods treat each task independently, overlooking the inherent reciprocity between them.
Moreover, these methods depend on paired training data, which is challenging to acquire, impeding the utilization of extensive unpaired data.
In this paper, we introduce MVSD, a mutual learning framework based on diffusion models. 
MVSD considers the two tasks symmetrically, exploiting the reciprocal relationship to facilitate learning from inverse tasks and overcome data scarcity.
Furthermore, we employ the diffusion model as foundational conditional converters to circumvent the training instability and over-smoothing drawbacks of conventional GAN architectures.
Specifically, MVSD employs two converters: one for VAM called reverberator and one for dereverberation called dereverberator.
The dereverberator judges whether the reverberation audio generated by reverberator sounds like being in the conditional visual scenario, and vice versa.
By forming a closed loop, these two converters can generate informative feedback signals to optimize the inverse tasks, even with easily acquired one-way unpaired data.
Extensive experiments on two standard benchmarks, \ie, SoundSpaces-Speech and Acoustic AVSpeech, exhibit that our framework can improve the performance of the reverberator and dereverberator and better match specified visual scenarios.
\end{abstract}
\section{Introduction}
\label{sec:intro}
Sound interacts with its environment, giving listeners a sense of objects and spatial imprints~\cite{DBLP:journals/taslp/ValimakiPSSA12}.
Reverberation is sound lingering in a space from surfaces reflecting sound waves~\cite{kuttruff2012room,gade2014acoustics}. 
Thus, reverberant sound, faithfully replicating real-world acoustics, is vital for realistic and immersive experiences in applications like augmented and virtual reality~\cite{WangWLY19,DBLP:conf/siggraph/MalpicaSGMBGM22,DBLP:conf/siggraph/KimGCPLW22,ChenWLLY23,li2023catr,DoraemonGPTyang,MS2SL24}.
Although reverberation can bestow a realistic sense of space, it may make speech content less intelligible~\cite{rennies2011prediction,kressner2018impact}.
In line with human perception, automatic speech recognition systems also suffer from lower accuracy when processing reverberant speeches~\cite{han2015learning,wu2017end,ernst2018speech}.
Therefore, dereverberation techniques~\cite{DBLP:journals/taslp/NakataniBKIDH20} can benefit applications such as teleconferencing, hearing aids, voice assistants, \etc.
\begin{figure}[t]
  \centering
  \includegraphics[width=1 \linewidth, trim=20bp 19bp 27bp 19bp, clip]{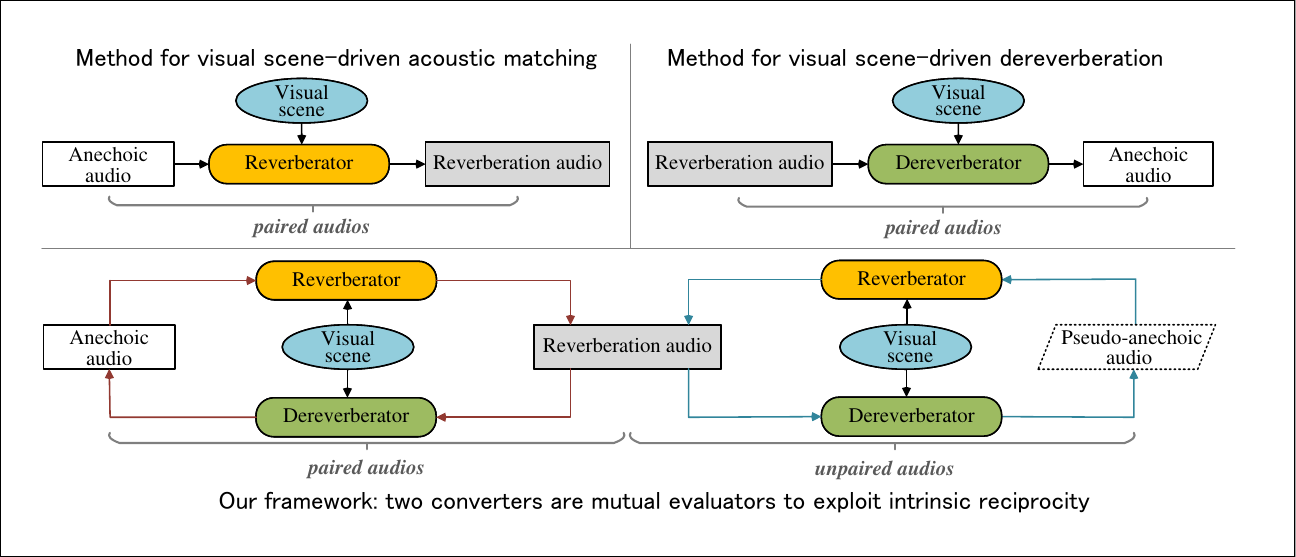}
\caption{There exists an inherent reciprocity between VAM and dereverberation. Unlike previous approaches that treat these two tasks independently, our framework simultaneously handles the both tasks. Forming a closed loop between the two converters can generate informative feedback signals to optimize the inverse tasks, even with easily acquired one-sided unpaired data~(\S\ref{sec:intro}).}
\label{fig:mutual learning}
\end{figure}
Existing works train VAM and dereverberation separately~\cite{DBLP:conf/cvpr/GaoG19,DBLP:conf/icassp/SuJF20,DBLP:conf/interspeech/FuYHPRL021,singh2021image2reverb,chen22vam,ChenSHG23}.
The traditional methods of acoustic matching primarily involve unraveling the spatial characteristics of sound through the examination of Room Impulse Responses (RIRs), which assess the propagation and variation of sound within a specific environment~\cite{funkhouser2004beam,bilbao2013modeling,cao2016interactive,murgai2017blind,savioja2020simulation,DBLP:conf/icassp/SuJF20}. 
Rather than estimating RIRs, VAM~\cite{chen22vam} directly achieves specified reverberation by employing images of the target environment and original audio clips.
For dereverberation, classical methodologies often encompass the application of signal processing and statistical techniques~\cite{nakatani2010speech,naylor2010speech},
recent advances highlight neural network-based approaches that learn transfer functions from reverberation to anechoic spectrograms~\cite{ernst2018speech,DBLP:journals/taslp/HanWWWMZ15,fu2019metricgan,DBLP:journals/taslp/WuLYL17}.
Nonetheless, optimizing each task individually fails to leverage the inherent reciprocity between the two tasks~(Fig.~\ref{fig:mutual learning}).
Moreover, training these methods usually requires extensive paired data.
Yet capturing large volumes of aligned anechoic and reverberant audio pairs in real-world scenarios is not feasible.
For VAM, the shortage of paired audio usually leads to average-style reverberation.
When it comes to dereverberation, the model struggles to produce highly `clean' audio in response to complex scenarios.
Thus, existing methods often face challenges in leveraging extensive unpaired audio due to the varying reverberation levels.

In this paper, we consider dereverberation as the inverse task of VAM, serving as an evaluator to provide feedback signals for VAM training, and vice versa. 
Specifically, given a visual environment~$\bm{v}$, an anechoic audio~$\bm{a}_c$, and a reverberant audio~$\bm{a}_r$, VAM reverberator $f_\theta(\bm{v}, \bm{a}_c) \rightarrow \hat{\bm{a}}_r$ maps the visual observation and anechoic audio into reverberant audio, while the dereverberator $g_\phi(\bm{v}, \bm{a}_r) \rightarrow \hat{\bm{a}}_c$ restores reverberant audio to anechoic audio conditioned on visual characteristics.
There exists a solid reciprocal relationship between the input and output spaces of $f_\theta$ and $g_\phi$. 
In this study, we delve into exploiting their intrinsic reciprocity to overwhelm the scarcity of parallel data. We propose a \underline{M}utual learning mechanism based on \underline{V}isual \underline{S}cene-driven \underline{D}iffusion (MVSD)~(Fig.~\ref{fig:mutual learning}).
In MVSD, two converters, namely reverberator and dereverberator, are employed and capable of learning from the symmetric tasks.
Taking VAM as an example, the reverberator, conditioned on the visual scene~$\bm{v}$, simulates environmental acoustic effects and converts anechoic audio~$\bm{a}_c$ to reverberant audio~$\hat{\bm{a}}_r$.
Since the output of one converter can be used as the input for another, the reverberator and dereverberator can act as mutual evaluators.
Concretely, in the primal task VAM, the reverberator generates reverberated audio~$\hat{\bm{a}}_r$ conditioned on the visual scene~$\bm{v}$ and anechoic audio $\bm{a}_c$.
Then the reverse converter $g_\phi$ takes $\hat{\bm{a}}_r$ as input and reconstructs the anechoic audio $\tilde{\bm{a}}_c$ within the symmetric dereverberation task.
Finally, the errors between $\tilde{\bm{a}}_c$ and $\bm{a}_c$ are used as feedback signals to optimize reverberator~$f_\theta$, and vice versa.
The training process of reverberator~$f_\theta$ and dereverberator~$g_\phi$ can form a closed loop, providing feedback for inverse tasks to enhance data efficiency.
When the dereverberator encounters a unpaired natural audio~$\bm{a}'_r$ with reverberation, it first eliminates the reverberation factors and creates a pseudo-anechoic audio~$\hat{\bm{a}}'_c$. Likewise, the reverberator regenerates $\tilde{\bm{a}}'_r$ based on $\hat{\bm{a}}'_c$ and visual observations~$\bm{v}'$.
Hence, MVSD allows these two converters to benefit from each other's training instances and can be extended to easily acquired unpaired audio samples.
For conditional generation, the architecture built on GANs is presently the prevailing choice~\cite{GoodfellowPMXWOCB14,DBLP:journals/corr/RadfordMC15,DBLP:conf/nips/ChenCDHSSA16,DBLP:journals/corr/MirzaO14,DBLP:conf/iclr/KarrasALL18,DBLP:conf/cvpr/KarrasLA19}.
However, the training of GAN may introduce potential risks of instability and over-smoothing.
Diffusion model~\cite{DBLP:conf/nips/HoJA20,DBLP:conf/iccv/ChoiKJGY21,DBLP:conf/iclr/MengHSSWZE22,DBLP:conf/cvpr/LugmayrDRYTG22,dhariwal2021diffusion,DBLP:conf/wacv/LiuPAZCHSRD23,DBLP:conf/cvpr/AvrahamiLF22,DBLP:conf/nips/SahariaCSLWDGLA22} recently show remarkable milestones in image generation, enabling the creation of high-quality images based on conditioning cues.
Some works introduce diffusion into audio generation, such as converting spectrograms into sound signals~\cite{KongPHZC21}, generating symbolic music~\cite{MittalEHS21},~\etc.
However, diffusion generation of specified reverberation styles under visual guidance remains underexplored.
To bridge this gap, we meticulously devise a visual scene-driven diffusion model to mitigate the computational overhead.
Specifically, the diffusion model for each task includes a visual scene encoder for extracting features to control reverberation style, and a controllable Unet that serves as the generator for producing the desired audio.
Additionally, cross-modal attention is adopted in selective blocks to establish correlations between visual cues and audio, reducing computational demands.

We spotlight the notable strengths of MVSD in visual-audio cross-modal style transfer.
MVSD effectively enhances the performances and consistently reports promising results on both tasks.
We achieve a remarkable reduction of $0.157$ in STFT-distance on the `Seen' test set of SoundSpaces-Speech~\cite{chen22vam} ($23.6\%$ relative performance).
Moreover, the utilization of unpaired audios~($17.3\%$ of the training data) can further boost the relative performance by $9.1\%$ in RTE for VAM.

\noindent To summarize, the main contributions of this paper are as follows:
\begin{itemize} [left=0em]
  \item We initially propose an end-to-end approach that leverages the reciprocity between VAM and dereverberation tasks to reduce reliance on paired data.
  \item We introduce a new and elegant mutual learning framework, MVSD, incorporating diffusion models and utilizing symmetrical tasks as evaluators to provide feedback signals to facilitate model training.
  \item We conduct a comprehensive evaluation of MVSD, demonstrating its superiority and confirming the potential of unpaired data in real-world applications.
\end{itemize}
\section{Related Work}
\label{sec:rw}
\textbf{Acoustic Matching.}
Acoustic matching involves modifying audio to simulate the sound in a given environment.
Schroeder~\etal~\cite{schroeder1961colorless} first propose the concept of reverberation and apply a series of percolators and delay lines to mimic environmental space characteristics.
There are two main methods for acquiring RIRs in the audio community~\cite{DBLP:conf/interspeech/MackDH20, gamper2018blind, murgai2017blind}. \textbf{(1)} Simulation techniques can be employed to produce RIRs when the geometry and material properties of the spatial environment are available~\cite{bilbao2013modeling, cao2016interactive, funkhouser2004beam}. \textbf{(2)} If detailed information is inaccessible, RIRs can be blindly estimated from audio captured in the room~\cite{murgai2017blind, DBLP:conf/icassp/SuJF20}.
RIRs are then employed to synthesize an auralized audio signal.
Both methods have weaknesses. The former requires exhaustive measurements of space that may be infeasible, while the latter may introduce some disturbances due to limited acoustic information.
Some recent works~\cite{kon2019estimation, singh2021image2reverb} attempt to approximate RIRs from an environmental image, necessitating paired image and impulse response training data.
Regrettably, these methods also require estimating the acoustic parameters from the recorded audio, which severely limits the application scopes.
Chen~\etal~\cite{chen22vam}~introduce VAM and utilize visual observation to simulate the target environment for generating reverberant audio.
However, VAM focuses on acoustic matching, neglecting the correlation and inherent consistency with the reverse dereverberation task.
In this paper, we harness the RGB image of specified environment for acoustic matching and utilize the reciprocity with dereverberation to improve the precision of reverberation simulations.

\noindent\textbf{Dereverberation.}
Due to the challenge of collecting both anechoic and reverberant audio simultaneously, acoustic dereverberation can enhance training data quality by minimizing reverberation disturbance~\cite{DBLP:conf/icassp/KoPPSK17, DBLP:journals/taslp/ZhaoWXZ20}.
The main stream dereverberation technologies utilize devices like microphone arrays to remove reverberation~\cite{DBLP:journals/tsp/MiyoshiK88}.
Deep learning techniques have also made great strides in reverberation removal~\cite{DBLP:journals/taslp/HanWWWMZ15, DBLP:journals/taslp/WuLYL17, DBLP:journals/taslp/ZhaoWW19}.
Tan~\etal~\cite{DBLP:journals/jstsp/TanXZYY20}~exploit the movement of the upper lip region to isolate interfering sounds, yet it does not intentionally eliminate reverberation based on visual scene understanding.
These methods either disregard or only partially take into account visual information.
Chen~\etal~\cite{ChenSHG23}~propose learning all the acoustics characteristics associated with indoor dereverberation.
Like acoustic matching, these unidirectional approaches neglect the reciprocal relationship between the two tasks, leading to an incomplete utilization of naturally recorded audio.
In contrast, MVSD demonstrates stronger dereverberation capabilities through the assistance of symmetric tasks. \\
\textbf{Mutual Learning.}
Mutual learning, originating from the field of language translation, aims to reduce dependence on data annotation~\cite{DBLP:conf/nips/HeXQWYLM16}.
This mechanism allows alternating between the two sides and enables the language model to train solely from one-sided data. 
The core idea of mutual learning involves establishing a dual-learning game between two agents, each agent is assigned an individual task.
In the primal task, mutual learning maps~$\bm{x}$ from primal domain to dual domain~$\bm{y}$, and then restore the original~$\bm{x}$ through the reverse mapping in dual task~\cite{DBLP:conf/acml/ZhaoXQXL20,WangLSGW22}.
Hence, mutual learning can produce two feedback signals without requiring parallel data: a style evaluation score indicating the likelihood that the synthesized audio matches the target style, and a reconstruction loss measuring the difference between the reconstructed audio and the original audio.
This mechanism alternates between agents, allowing the generator to train from only one-way data~\cite{DBLP:conf/iccv/YiZTG17,DBLP:conf/nips/LuKYPB17,DBLP:conf/cvpr/ShahCRP19,DBLP:conf/acml/ZhaoXQXL20,DBLP:conf/nips/XieDHL020,DBLP:conf/cvpr/ZhangXHL18}.
We are the first to investigate the duality of VAM and dereverberation.
These two tasks are trained together in a mutual learning framework and provide mutual reinforcement signals based on the structural symmetry, even for unpaired samples.

\noindent\textbf{Condition-guided Generation.}
In recent years, there have been significant advancements in the field of conditional generation~\cite{DBLP:conf/iccv/HuS21,DBLP:conf/cvpr/SinghHGCGRK22,DBLP:conf/icml/WangYMLBLMZZY22,DBLP:conf/icml/RadfordKHRGASAM21}.
Diffusion models have demonstrated impressive results  in various generative tasks due to their superior visual quality and training stability~\cite{DBLP:conf/nips/HoJA20,DBLP:conf/iccv/ChoiKJGY21,DBLP:conf/cvpr/LugmayrDRYTG22,DBLP:conf/wacv/LiuPAZCHSRD23,DBLP:conf/cvpr/AvrahamiLF22,DBLP:conf/nips/SahariaCSLWDGLA22,nichol2021improved, dhariwal2021diffusion,KongPHZC21, MittalEHS21}.
The diffusion probability model~\cite{DBLP:conf/icml/Sohl-DicksteinW15} is based on a Markov chain, proceeding through finite steps in two opposing directions: one transition moves from the data distribution to noise, and the other transitions back from noise to the data distribution.
Ho~\etal~\cite{DBLP:conf/nips/HoJA20} introduce the variational lower bound objective, which is subsequently improved in~\cite{nichol2021improved} to obtain higher log-likelihood scores.
In this study, we regard audio spectrograms as images and elegantly employ two diffusion-based generators for controllable reverberation style transfer.
\section{Methodology}
\label{sec:method}
\vspace{-3pt}
We propose a mutual learning framework MVSD to leverage feedback signals from symmetrical tasks to promote model training and better exploit unpaired data.
It involves two tasks: a primal task VAM~\cite{chen22vam} that employs the reverberator $f_\theta$ to convert an anechoic audio~$\bm{a}_c$ into a reverberated audio~$\hat{\bm{a}}_r$, which is aurally recorded in the specified environment.
In the dual task, dereverberator $g_\phi$ removes the reverberant characteristics in~$\bm{a}_r$, which is similar to its anechoic counterpart~$\bm{a}_c$.
Here, $f_\theta$ and $g_\phi$ are jointly trained in an end-to-end mutual learning framework MVSD~(\S\ref{sec:MVD}).
Furthermore, we employ visual scene-driven diffusion models as foundational conditional converters~$f_\theta$~and~$g_\phi$ to achieve stable training and accurate reverberation style transfer~(\S\ref{sec:DD}).

\noindent\textbf{Reverberator.}
Consider paired data distributions: $\!\bm{A}_c\!=\!\{\bm{a}_c^{(1)},\!\bm{a}_c^{(2)},\!...,\!\bm{a}_c^{(n)}\}$ and $\bm{A}_r=\{\bm{a}_r^{(1)},\bm{a}_r^{(2)},...,\bm{a}_r^{(n)}\}$, representing anechoic and reverberant audio, respectively.
The set of visual scenes $\bm{V}=\{\bm{v}^{(1)},\bm{v}^{(2)},...,\bm{v}^{(n)}\}$ corresponds to the audio set of $\bm{A}_r$.
The goal of VAM is to convert the anechoic audio $\bm{a}_c$ with condition $\bm{v}$ to its reverberant counterpart~$\bm{a}_r$, \ie, to estimate the conditional distribution $f_\theta(\bm{a}_r|\bm{a}_c;\bm{v})$.
Based on diffusion models, we encode $\bm{a}_c$ into content features and switch the reverberation style to the visual environment~$\bm{v}$.

\noindent\textbf{Dereverberator.} 
Contrary to VAM, the goal of the dereverberation task is to eliminate reverberation factors and enhance the intelligibility of audio content.
Correspondingly, the dereverberator~$g_\phi$ based on VSD calculates the anechoic distribution~$g_\phi(\bm{a}_c|\bm{a}_r;\bm{v})$ under a given scene~$\bm{v}$.
\begin{figure*}[t]
  \centering
     \includegraphics[width=1  \linewidth, trim=20bp 22bp 25bp 22bp, clip]{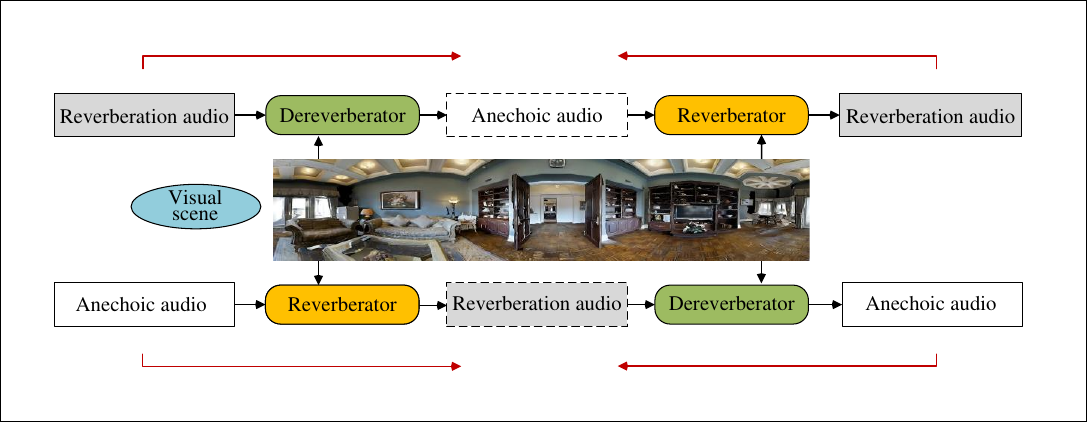}
     \put(-315,12){\scalebox{.80}{$\bm{a}_c$}}
     \put(-257,11){\scalebox{.80}{$f_\theta(\bm{v},\bm{a}_c)$}}
     \put(-177,11){\scalebox{.80}{$\hat{\bm{a}}_r$}}
     \put(-119,11){\scalebox{.80}{$g_\phi(\bm{v},\hat{\bm{a}}_r)$}}
     \put(-38,11){\scalebox{.80}{$\tilde{\bm{a}}_c$}}
     \put(-315,101){\scalebox{.80}{$\bm{a}'_r$}}
     \put(-258,101){\scalebox{.80}{$g_\phi(\bm{v}',\bm{a}'_r)$}}
     \put(-179,101){\scalebox{.80}{$\hat{\bm{a}}'_c$}}
     \put(-121,101){\scalebox{.80}{$f_\theta(\bm{v}',\hat{\bm{a}}'_c)$}}
     \put(-38,101){\scalebox{.80}{$\tilde{\bm{a}}'_r$}}
     \put(-198,1){\scalebox{.75}{$\Delta^{\bm{a}_c}_{\bm{v}}=\|\tilde{\bm{a}}_c, \bm{a}_c\|_1$}}
     \put(-198,111.5){\scalebox{.75}{$\Delta^{\bm{a}'_r}_{\bm{v}'}=\|{\tilde{\bm{a}}'_r}, {\bm{a}'_r}\|_1$}}
\caption{The overview of MVSD. 
The output of a converter can serve as pseudo-input for the reverse task, providing an intermediate transition.
Concretely, the reverberator $f_\theta$ and dereverberator $g_\phi$ can generate feedback signals $\mathcal{L}_{m}$~(Eq.~\ref{eq:6}) for mutual optimization of training, even with one-way unpaired data~($\bm{a}'_r, \bm{v}'$)~(\S\ref{sec:MVD}).
}
\label{fig:framework}
\vspace{-8pt}
\end{figure*}
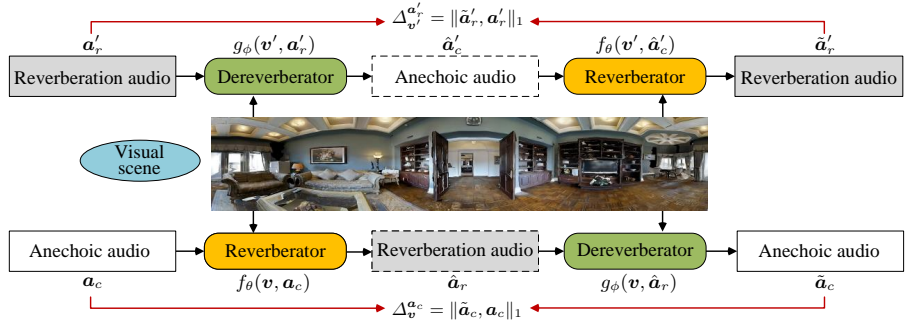
\subsection{Mutual Learning}\label{sec:MVD}
We jointly learn the VAM and dereverberation tasks~(Fig.~\ref{fig:framework}):
the reverberator~$f_\theta$ and dereverberator~$g_\phi$ can mutually benefit from each other.
Suppose we have two (vanilla) converters that can map anechoic audio to a specified reverberation style and vice versa. 
Our goal is to simultaneously improve the style accuracy of the VAM task and the content intelligibility of the dereverberation task by employing paired and unidirectional non-paired data. 
To achieve this, we leverage the reciprocity between these two tasks, wherein the input-output spaces of VAM and dereverberation exhibit a strong correlation and can interchangeably act as the input and output for each other.  
Starting from either task, we first convert it forward to another audio, then transfer it backward to the original audio. 
By evaluating  the results of this two-hop transfer process, we can gauge the quality of both converters and optimize them accordingly. 
Namely, dereverberator~$g_\phi$ is employed to evaluate the quality of $\hat{\bm{a}}_r$ generated by $f_\theta$ and sends back an error singal $\triangle(\tilde{\bm{a}}_c, \bm{a}_c)$ to $f_\theta$, and vice versa.
This process can be iterated many rounds until both converters converge.
Please note that in MVSD, $\bm{a}_r$ and $\bm{a}_c$ are not necessarily aligned and may even not have a typical relationship.

We~denote a labeled collection as $\mathcal{D}\!=\!\{(\bm{v}^n, \bm{a}_c^n,$ $\bm{a}_r^n)\}_{n=1}^N$, which consists of $N$ aligned tuples of anechoic and reverberant audio. 
Given a triplet~$\langle \bm{v},\bm{a}_c, \bm{a}_r \rangle$, where  $\bm{v}$, $\bm{a}_c$, $\bm{a}_r$ are sets of \textit{environmental spaces}, \textit{anechoic} and \textit{target audios}.
Our goal is to uncover the bi-directional relationship between the $\bm{a}_c$ and $\bm{a}_r$.
For the primal process starting from VAM, denote $\hat{\bm{a}}_r$ as the mid-transition output.
Firstly, we obtain a reverberated audio $\hat{\bm{a}}_r$ through the reverberator $f_\theta(\bm{v}, \bm{a}_c)$.
Then, the dereverberator $g_\phi$ translates $\hat{\bm{a}}_r$~to~$\tilde{\bm{a}}_c$ by mapping $g_\phi(\bm{v}, \hat{\bm{a}}_r)$.
The~$\tilde{\bm{a}}_c$ is expected to be consistent with $\bm{a}_c$ in audio clarity, \ie, achieving a small cycle-consistency error~$\Delta^{\bm{a}_c}_{\bm{v}}$.
Similarly, for dereverberator~$g_\phi$, we have~$\tilde{\bm{a}}_r=f_\theta(\bm{v}, \hat{\bm{a}}_c)$ and $\tilde{\bm{a}}_r$~should have a reverberation effect akin to $\bm{a}_r$ in auditory perception. 
Likewise, $\Delta^{\bm{a}_r}_{\bm{v}}$~can be employed to evaluate the discrepancies between $\bm{a}_r$ and $\tilde{\bm{a}}_r$.
Finally, the errors~$\Delta^{\bm{a}_c}_{\bm{v}}$ and~$\Delta^{\bm{a}_r}_{\bm{v}}$ can be specified as two reconstruction losses, which are minimized for the model training. 
Prior researches~\cite{LarsenSLW16,DBLP:conf/nips/HeXQWYLM16} on conditional image synthesis suggest that $L1$ distance, unlike $L2$, can reduce blurriness. Hence, we employ $L1$ distance to assess the feedback errors:
\begin{equation}\small\label{eq:3}
\begin{aligned}
\Delta^{\bm{a}_r}_{\bm{v}} = \|\tilde{\bm{a}}_r, \bm{a}_r\|_1 =\|f_\theta(\bm{v}, g_\phi(\bm{v}, \bm{a}_r)) - \bm{a}_r\|_1;\\
\Delta^{\bm{a}_c}_{\bm{v}} = \|\tilde{\bm{a}}_c, \bm{a}_c\|_1 = \|g_\phi(\bm{v}, f_\theta(\bm{v}, \bm{a}_c)) - \bm{a}_c\|_1.
\end{aligned}
\end{equation}
In real-world scenarios, the challenge of capturing parallel data arises from the difficulty of simultaneously recording sound at the source and listener locations.
This obstacle is mitigated in our approach, as it does not necessitate aligned anechoic and reverberant pairs~$(\bm{a}_c, \bm{a}_r)$ for the errors $\Delta^{\bm{a}_r}_{\bm{v}}$ and $\Delta^{\bm{a}_c}_{\bm{v}}$, As a result, Eq.~\ref{eq:3} can be effectively applied to one-way unpaired audios.
As in common practice~\cite{fried2018speaker,tan2019learning,ke2019tactical}, we build two unlabeled collections: $\mathcal{U}=\{({\bm{v}^m}', {\bm{a}_r^m}')\}_{m=1}^M$~for audios with natural reverberation, and $\mathcal{C}=\{{\bm{a}_c^k}''\}_{k=1}^K$ with only anechoic audio.
We obtain $\mathcal{U}$ by sampling natural audios ${\bm{a}'_r}$ from existing environments ${\bm{v}}'$, which lack corresponding anechoic audios. Similarly, we create collection~$\mathcal{C}$ by filtering anechoic audios~\cite{ChenSHG23} ${\bm{a}''_c}$ from an open-source dataset~\cite{DBLP:conf/icassp/PanayotovCPK15}, which do not have matching reverberated audios and visual images.
For unpaired natural audios~$\mathcal{U}$, we first generate intermediate output~$\hat{\bm{a}}'_c$ using the dereverberator~$f_\theta(\bm{a}'_r, \bm{v}')$, followed by reconstructing~$\tilde{\bm{a}}'_r$ based on~$\hat{\bm{a}}'_c$ and scene~$\bm{v}'$, \ie,~$g_\phi({\bm{v}}', {\bm{a}'_r})$, and computing error~$\Delta^{\bm{a}'_r}_{\bm{v}'}$ against the original input~$\bm{a}'_r$.
We can derive the formula:
\begin{equation}\small\label{eq:4}
\begin{aligned}
\Delta^{\bm{a}'_r}_{\bm{v}'} = \|{\tilde{\bm{a}}'_r}, {\bm{a}'_r}\|_1 = \|f_\theta({\bm{v}}', g_\phi({\bm{v}}', {\bm{a}'_r}) - {\bm{a}'_r}\|_1.
\end{aligned}
\end{equation}
\noindent As for unpaired anechoic audios~$\mathcal{C}$, since there are rarely accompanying visual scene images when recording audio, we randomly sample an image~$\bm{v}''$ from~$\mathcal{U}$ to simulate the specified environment, and the formulate is as:
\begin{equation}\small\label{eq:5}
\begin{aligned}
\Delta^{\bm{a}''_c}_{\bm{v}''} = \|{\tilde{\bm{a}}''_c}, {\bm{a}''_c}\|_1 = \|g_\phi(\bm{v}'', f_\theta(\bm{v}'', {\bm{a}''_c})) - {\bm{a}''_c}\|_1.
\end{aligned}
\end{equation}
\noindent Our training process utilizes paired data, complemented by unpaired natural and anechoic audios. Consequently, our mutual learning loss is defined as:
\begin{equation}\small\label{eq:6}
\begin{aligned}
\mathcal{L}_{m}\!=&\!\frac{1}{N}\!\sum\nolimits_{({\bm{v}},\bm{a}_r,\bm{a}_c)~\!\in~\!\mathcal{D}}\!(\Delta^{\bm{a}_c}_{\bm{v}}\!+\!\Delta^{\bm{a}_r}_{\bm{v}})\!\!+\!\!
\frac{1}{M}\!\sum\nolimits_{({\bm{v}'}, {\bm{a}'_r})~\!\in~\!\mathcal{U}}\!\!\triangle^{\bm{a}'_r}_{\bm{v}'}\!\!+\!\!\frac{1}{K}\!\sum\nolimits_{({\bm{v}''}, {\bm{a}''_c})~\!\in~\!\mathcal{C}}\!\!\triangle^{\bm{a}''_c}_{\bm{v}''}.
\end{aligned}
\end{equation}
\noindent During training, $\mathcal{L}_{m}$ is applied only for predictions and backpropagation at time step $t$ of the diffusion model. Hence, MVSD does not significantly increase the training time compared to training the two tasks separately.

\noindent\textbf{Remark.}
MVSD consists of two main concepts: First, an ideal reverberator should be able to adapt audio to any visual environment, and 
a dereverberator is also effective at removing disturbances that affect speech intelligibility.
Therefore, we investigate VAM and dereverberation in a unified learning framework, allowing the converters to better exploit the cross-modal and cross-task correlations.
Second, the addition of unpaired data can boost model performance, and paired data guides the reverberator and dereverberator converge to the target distribution, preventing extreme domain deviation from the unpaired data.
\subsection{Visual Scene-driven Diffusion} \label{sec:DD}
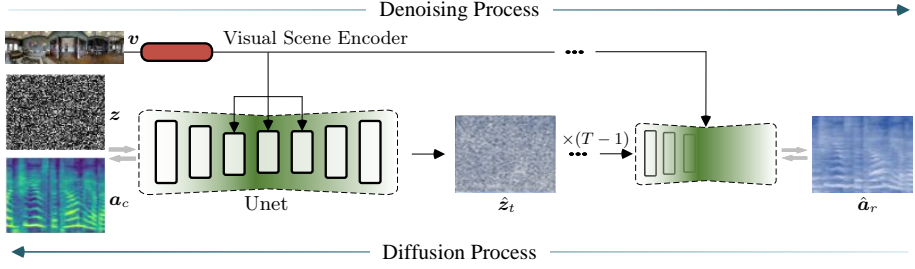
\begin{figure*}[t]
  \centering
     \includegraphics[width=1  \linewidth, trim=20bp 20bp 20bp 22bp, clip]{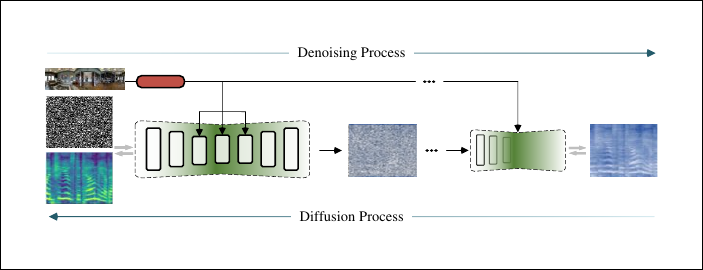}
     \put(-299,85){\scalebox{.80}{$\bm{v}$}}
     \put(-306,56){\scalebox{.80}{$\bm{z}$}}
     \put(-306,24){\scalebox{.80}{$\bm{a}_c$}}
     \put(-159,22){\scalebox{.80}{$\hat{\bm{z}}_t$}}
     \put(-23,22){\scalebox{.80}{$\hat{\bm{a}}_r$}}
     \put(-255,22){\scalebox{.85}{Unet}}
     \put(-263,85){\scalebox{.80}{Visual Scene Encoder}}
     \put(-135,47){\scalebox{.70}{$\times(T-1)$}} 
\caption{The diffusion and denoising processes of VSD. 
Taking VAM as an example, MVSD converts anechoic audio~$\bm{a}_c$ into reverberant audio~$\hat{\bm{a}}_r$ that aligns with the acoustics of the visual scene~$\bm{v}$~(\S\ref{sec:DD}).
}
\label{fig:vsd}
\vspace{-8pt}
\end{figure*}
In MVSD, the reverberator and dereverberator share a similar model structure. We introduce visual scene-driven diffusion (VSD) with the reverberator~$f_\theta$ as an example.
The diffusion model employs a $T$-step iterative denoising process to transform Gaussian noise into the desired data distribution~\cite{DBLP:conf/icml/Sohl-DicksteinW15,DBLP:conf/nips/HoJA20,DBLP:conf/nips/SahariaCSLWDGLA22}.
By introducing prompt conditions such as class labels and text~\cite{dhariwal2021diffusion,DBLP:conf/icml/NicholDRSMMSC22}, the generated content can be controlled precisely. 
In MVSD, visual scene embeddings are employed as control conditions to guide the generation of reverberator~$f_\theta$ and dereverberator~$g_\phi$.
In particular, the diffusion process follows a Markov chain, progressively adding noise to the input spectrogram~$\bm{x}_0$~(sampled from the real distribution $q(\bm{x})$ until it evolves into white Gaussian noise~$\mathcal{N}(0, 1)$. 
At each step~$t$, the spectrogram $\bm{x}_t$, following the distribution $q(\bm{x}_t|\bm{x}_{t-1})$, is derived by 
the pre-defined variance $\beta_t$ scaled with $\sqrt{1-\beta_t}$:
\begin{equation}\small\label{eq:1}
    q(\bm{x_t}|\bm{x}_{t-1}) = \mathcal{N}(\bm{x}_t;\bm{z}_t); \quad \bm{z}_t\sim\mathcal{N}(\sqrt{1-\beta_t}\bm{z}_{t-1}, \beta_t\textbf{I}).
\end{equation}
The denoising process attempts to restore the original spectrogram~$\bm{x}_0$ from the noisy data~$\bm{x}_T$ by removing the noise introduced in the forward diffusion process.
The prediction $q(\bm{x}_{t-1}|\bm{x}_t)$ at step $t-1$ is approximated by a parameterized model~$p$~(\eg, a neural network), involving the estimation of $\mu(\bm{z}_t, t)$ and $\sigma(\bm{z}_t, t)$ from a Gaussian distribution.
By employing the reverse process across all time steps, we can transition from $\bm{x}_T$ back to the initial spectrogram~$\bm{x}_0$:
\begin{align}\small\label{eq:2}
    \!\!\!p(\bm{x}_{0:T}) &\!=\! p(\bm{x}_T)\prod\nolimits_{t=1}^Tp(\bm{x}_{t-1}|\bm{x}_t) \notag \\
                           &\!=\!p(\bm{x}_T)\prod\nolimits_{t=1}^T\mathcal{N}(\bm{x}_{t-1};\mu(\bm{x}_t, t), \sigma(\bm{x}_t, t)).
\end{align}
\noindent\textbf{Visual Scene Encoder.}
We apply an embedding with $256$ dimensions to represent visual scenes, extracted by a pre-trained ResNet-18~\cite{he2016deep} encoder. 
Then, the embedding serves as the condition to guide the generation of diffusion models.

\noindent\textbf{Controllable Unet.}
We meticulously design an controllable Unet for predicting~$\bm{x}_t$ of diffusion~(Fig.~\ref{fig:vsd}).
Controllable Unet is composed of multiple stages with attention blocks~\cite{DBLP:conf/nips/SahariaCSLWDGLA22}, \ie, self-attention and cross-attention.
Self-attention allows a model to weigh the importance of different parts within the same element.
Cross-attention, similar to self-attention, targets relationships across different components.
We employ a classic encoder-decoder with a symmetric design, where each part incorporating $3$ attention blocks.
The encoder progressively reduces the resolution of the feature map, and then the decoder gradually increases it to align with the size of the original spectrogram.
In the self-attention block, we utilize the downsampling method in~\cite{DBLP:conf/pkdd/SunkaraL22} with a stride of $4$ to rapidly decrease the size of feature maps. 
The downsampling utilizes dilated convolutions and attention to increase the receptive field without reducing spatial dimensions.
Cross-modal attention is selectively employed to the third encoder block and the first decoder block, mitigating computational overhead.
Both VAM and dereverberation need to preserve the linguistic information in the audios.
Therefore, we concatenate source spectrogram with the noise~$\bm{z}_0$ as the content input for the controllable Unet. 
Please refer to the supplementary material for details. 
\subsection{Training Objective} \label{sec:TO}
For training the diffusion model, we employ the simplified objective~\cite{DBLP:conf/nips/HoJA20}:
\begin{equation}\small\label{eq:7}
\begin{aligned}
\mathcal{L}_{d}=\mathbb{E}_{{\bm{x}_0},t,\bm{z}}[\|\bm{z}-\hat{\bm{z}}(\sqrt{\overline{\alpha}_t}\bm{x}_0 + \sqrt{1-\overline{\alpha}_t}\bm{z},t)\|_2],
\end{aligned}
\end{equation}
where $\alpha_t$ in diffusion models is a scaling factor that modulates the noise level at each time step $t$.
VSD can predict the noise $\hat{\bm{z}}_t$ and use it to iteratively refine the denoising process.
With the reparameterization trick, a method for differentiable sampling~\cite{KingmaW13}, we can represent the estimation of $\hat{\bm{x}}_{0}$:
\begin{equation}\small\label{eq:9}
\begin{aligned}
\hat{\bm{x}}_{0}=\frac{1}{\sqrt{\overline{\alpha}_t}}(\bm{x}_t - \sqrt{1-\overline{\alpha}_t}\hat{\bm{z}}_t).
\end{aligned}
\end{equation}
Moreover, we introduce a style loss~$\mathcal{L}_{sty}$~(Eq.~\ref{eq:10}) to make the generated audios with the environmental characteristics.
Taking VAM task as an example, during training, the Unet predicts the noise~$\hat{\bm{z}}_t$ at time step $t$. 
Then, $\hat{\bm{z}}_t$ can be used to gradually derive the predicted original spectrogram~$\hat{\bm{x}}_r$ at step~$0$~(Eq.~\ref{eq:9}). 
Here, we do not explicitly extract the stylistic features of the~$\bm{a}_r$~and~$\hat{\bm{a}}_r$; instead, we directly employ $\mathcal{L}_1$ loss to regularize style consistency:
\begin{equation}\small\label{eq:10}
\begin{aligned}
\mathcal{L}_{sty} = \|\hat{\bm{a}}_r - \bm{a}_r\|_1 + \|\hat{\bm{a}}_c - \bm{a}_c\|_1.
\end{aligned}
\end{equation}
We learn models $f_\theta$~and~$g_\phi$ by minimizing the combination of the diffusion loss, the style loss  and the mutual learning regularization term. In summary, the overall training objective is given as:
\begin{equation}\small\label{eq:11}
    \mathcal{L}_{total} = \mathcal{L}_{d} + \mathcal{L}_{m} + \mathcal{L}_{sty}.
\end{equation}
\subsection{Implementation Details} \label{ID}
\noindent\textbf{Training.}
In MVSD, converters and visual scene encoder are trained separately.
We adopt the loss function in~\cite{DBLP:conf/nips/KhoslaTWSTIMLK20} to train the visual scene encoder.
The mutual learning is integrated into each mini-batch update, spanning the entire training process for the two tasks.
Training starts with supervised data, with unsupervised data progressively merged for optimization. This stepwise strategy can preserve model stability.
At each iteration, we compute the predictions of both converters and update their parameters based on the feedback from the symmetrical models.
In practice, we first perform supervised training and conduct the loop of mutual learning~(Alg.~\ref{algo:ML}).
Besides minimizing the cycle-consistent loss $\mathcal{L}_{m\!}$ (Eq.~\ref{eq:6}), our MVSD framework is learnt with the diffusion objectives for VAM and dereverberation, over the labeled data~$\mathcal{D}$.
Finally,$\!$ we$\!$ receive$\!$ a$\!$ prepared$\!$ model$\!$ when$\!$ MVSD$\!$ converges$\!$ on$\!$ all$\!$ training$\!$ data. 

\vspace{8pt}
\begin{algorithm}[H] \small 

    \caption{Mutual learning with visual scene-driven diffusion.}
    \label{algo:ML}
    \KwIn{Labeled set~$\mathcal{D}$, unpaired sets~$\mathcal{U}$~and~$\mathcal{C}$, reverberator $f_\theta$ and dereverberator~$g_\phi$.}
    \textbf{\textit{Repeat:}}
    ~~~~Sample a mini-batch of paired tuples~$\langle\bm{a}_c, \bm{v}, \bm{a}_r\rangle$;\\
    ~~~~Generate random Gaussian noise $\bm{z}_c$ and $\bm{z}_r$ for the converters; \\
    ~~~~Execute the diffusion processes of reverberator $f_\theta$ and dereverberator~$g_\phi$; \\
    ~~~~Calculate the training objective $\mathcal{L}_{total}$~(Eq.~\ref{eq:11}); \\
    ~~~~Update the parameters of $\theta$ and $\phi$: $\theta\leftarrow\theta-\gamma\nabla_\theta\mathcal{L}(\theta)$, $\phi\leftarrow\phi-\gamma\nabla_\phi\mathcal{L}(\phi)$;\\
    ~~~~Introduce unpaired data and continue training when the epoch exceeds $100$; \\ 
    \textbf{\textit{Until:}} Convergence
\end{algorithm}
\vspace{8pt}

\noindent\textbf{Inference.}
The inference of each task follows the sampling process of the diffusion model.
Take VAM as an example: First, a noise spectrogram is randomly generated and concatenated with a anechoic test spectrogram.
Next, at each step $t$ of the denoising process, the controllable Unet synthesizes the intermediate spectrogram conditioned on visual features.
We report the average of ten experiments as the evaluation criterions.

\noindent\textbf{Reproductibility.}
Our model is implemented in PyTorch and trained using two NVIDIA Tesla V100 GPUs. Training MVSD from scratch takes approximately 144 hours. The average inference time is $1.09$ seconds.
We set FFT size, hop size and mel scale for audio processing to $1024$, $256$ and $128$, respectively.
We then truncate the mel-spectrogram to a width of $128$, resulting in a spectrogram size of $128\times128$.
We utilize a pre-trained BigVGAN~\cite{DBLP:journals/corr/abs-2206-04658} as the vocoder. 
\section{Experiments}
\label{sec:exp}
\textbf{Dataset.}
We conduct experiments on two datasets~\cite{chen22vam}: \textit{SoundSpaces-Speech} and \textit{Acoustic AVSpeech} datasets.
The former employs a simulated environment~\cite{DBLP:conf/eccv/ChenJSGAIRG20} to generate reverberation audio, is perfectly aligned paired audio and accurate ground truth. Regardless, there has a realism gap.
Finally, the dataset is split into \textit{train/val/test} sets with $28,853$, $1,441$, and $1,489$ samples, respectively.
Acoustic AVSpeech is a subset of AVSpeech dataset~\cite{ephrat2018looking}.
It offers more realism but poses evaluation challenges due to lacking corresponding reverberant audios.
Acoustic AVSpeech contains $113k/3k/3k$ video clips for the training, validation, and test splits, respectively.
For the unpaired data, we randomly sample $5,000$ natural audios with video in AVSpeech dataset~\cite{ephrat2018looking} and $5k$ anechoic audio in Librispeech~\cite{DBLP:conf/icassp/PanayotovCPK15}, with no overlap with the training sets.
We apply `Seen' and `Unseen' to denote whether visual scenes are encountered during training.

\noindent\textbf{Evaluation Metrics.}
For VAM task, following~\cite{chen22vam}, we employ STFT-distance to measure deviation from the ground truth, Reverberation Time 60 error~(RTE) for room acoustics, and Mean Opinion Score Error~(MOSE) for speech quality evaluation.
For dereverberation task, as in~\cite{ChenSHG23}, we adopt Perceptual Evaluation of Speech Quality~(PESQ)~\cite{DBLP:conf/icassp/RixBHH01}, Word Error Rate~(WER) and Equal Error Rate~(EER) to assess the aspects of audio quality, content precision, and speaker verification error, respectively.
The evaluation is conducted on the anechoic version of LibriSpeech test set~\cite{DBLP:conf/icassp/PanayotovCPK15, ChenSHG23}.
\subsection{Performance on VAM} \label{PVAM}
\begin{figure*}[b]
\vspace{-8pt}
\centering
  \includegraphics[width=1  \linewidth, trim=20bp 22bp 20bp 22bp, clip]{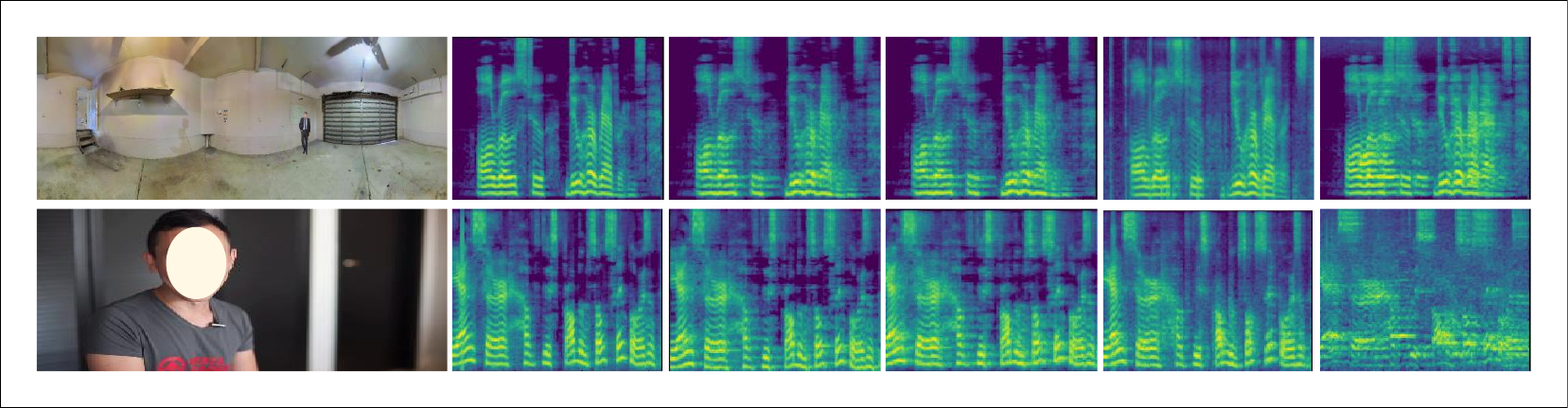}
  \put(-318,78){\scalebox{.80}{\texttt{Visual scene}}}
  \put(-251,78){\scalebox{.75}{\texttt{Anechoic audio}}}
  \put(-180,78){\scalebox{.75}{\texttt{GT}}}
  \put(-143,78){\scalebox{.75}{\texttt{MVSD\,(Ours)}}}
  \put(-91,78){\scalebox{.75}{\texttt{AViTAR$\!\!\!\!$\cite{chen22vam}}}}
  \put(-54,78){\scalebox{.75}{\texttt{Image2Reverb$\!\!\!\!$\cite{singh2021image2reverb}}}}
  \caption{Visualization results for VAM task on the SoundSpaces-Speech (top) and AVSpeech datasets (bottom)~\cite{chen22vam}~(\S\ref{PVAM}). }
 \label{fig:VisReverb}
\end{figure*}

\begin{table*}[t]
	\centering
    \caption{Quantitative results on \textit{SoundSpaces-Speech} and \textit{Acoustic AVSpeech}~\cite{chen22vam} (\S\ref{PVAM}).} 
	\setlength\tabcolsep{0.8pt}
	\renewcommand\arraystretch{1.3}
	\resizebox{1\linewidth}{!}{
		\begin{tabular}{l||c|c|c|c|c|c||c|c|c|c}
			\thickhline
            \thickhline
            \rowcolor{mygray}
            & \multicolumn{6}{c||}{\textit{SoundSpaces-Speech}}                                                                                                       & \multicolumn{4}{c}{\textit{Acoustic AVSpeech}}                                                 \\
            \rowcolor{mygray}
            Method & \multicolumn{3}{c|}{\textit{Seen}}                                                   & \multicolumn{3}{c||}{\textit{Unseen}}                            & \multicolumn{2}{c|}{\textit{Seen}}                       & \multicolumn{2}{c}{\textit{Unseen}} \\
            \rowcolor{mygray}
           & \multicolumn{1}{c|}{STFT $\downarrow$} & \multicolumn{1}{c|}{RTE (s)  $\downarrow$} & \multicolumn{1}{c|}{MOSE  $\downarrow$} & \multicolumn{1}{c|}{STFT  $\downarrow$} & \multicolumn{1}{c|}{RTE (s)  $\downarrow$} & MOSE  $\downarrow$ & \multicolumn{1}{c|}{RTE (s)  $\downarrow$} & \multicolumn{1}{c|}{MOSE  $\downarrow$} & \multicolumn{1}{c|}{RTE (s)  $\downarrow$} & MOSE  $\downarrow$ \\
			\hline
			\hline
			Input audio          & 1.192 & 0.331 & 0.617 & 1.206 & 0.356 & 0.611 & 0.387 & 0.658 & 0.392 & 0.634 \\
            AEE~\cite{DBLP:conf/icassp/SuJF20}  & 2.746 & 0.319 & 0.571 & - & - & - &  -   &   -   &  -    &  - \\
            Image2Reverb~\cite{singh2021image2reverb}         & 2.538 & 0.293 & 0.508 & 2.318 & 0.317  & 0.518 &  -   &   -   &  -    &  - \\
            AV U-Net~\cite{DBLP:conf/cvpr/GaoG19}             & 0.638 & 0.095 & 0.353 & 0.658 & 0.118 & 0.367 & 0.156 & 0.570 & 0.188 & 0.540 \\
            AViTAR~\cite{chen22vam}               & 0.665 & 0.034 & 0.161 & 0.822 & 0.062 & 0.195 & 0.144 & 0.481 & 0.183 & 0.453 \\
            \cdashline{1-11}[1pt/1pt]
            MVSD \textit{w/o} visual scene       & 0.691 & 0.188 & 0.156 & 0.803 & 0.155 & 0.194 & 0.137 & 0.526 & 0.171 & 0.474 \\
            MVSD \textit{w/o} unpaired data       & 0.573 & 0.033 & 0.148 & 0.736 & 0.055 & 0.184 & 0.131 & 0.427 & 0.159 & 0.394 \\
            \textbf{MVSD}                         & \bfseries 0.508 & \bfseries 0.030 \bfseries & \bfseries 0.142 & \bfseries 0.637 & \bfseries 0.051
                                        & \bfseries 0.178 & \bfseries 0.112 & \bfseries 0.392 & \bfseries 0.148 & \bfseries 0.379 \\
			\hline
		\end{tabular}}
		\label{tab:quan-vam}
    \vspace*{-10pt}
\end{table*}
As shown in Table~\ref{tab:quan-vam}, MVSD achieves a notable absolute boost of $0.157$ STFT-distance ($23.6\%$ relative improvement), $0.004$ RTE ($11.8\%$ relative improvement), and $0.019$ MOSE ($11.8\%$ relative improvement) in the `Seen' split of SoundSpaces-Speech dataset compared with SOTA method.
There is also a similar improvement in Acoustic AVSpeech dataset.
We can see that MVSD exhibits outstanding strengths in all three assessed aspects: preserving source audio content better, getting in more precise signal attenuation and more consistent quality with the target audio.
MVSD has the capability to infer and extract relevant factors that influence reverberation from target images, even in never-before-seen scenes.
It should be noted that blind reverberator~\cite{DBLP:conf/icassp/SuJF20}, a traditional acoustic method, needs reference audio, making it unsuitable for scenarios `Unseen' of SoundSpaces~(no reference audio) and AVSpeech, as reported in~\cite{chen22vam}.
Fig.~\ref{fig:VisReverb} showcases the visual comparisons of different methods for the VAM task on the SoundSpaces-Speech and AVSpeech datasets, respectively, highlighting the superiority of MVSD.
\begin{figure*}[t]
        \begin{minipage}[t]{0.59\textwidth}
            \makeatletter\def\@captype{table}\captionsetup{width=.92\linewidth}
            \centering \small
            \caption{Quantitative dereverberation results on SoundSpaces-Speech\!~\cite{chen22vam}\!~(\S\ref{PD}).\!\!\!\!\!\!} 
            \resizebox{1.\linewidth}{!}{
                \setlength\tabcolsep{1pt}
                \renewcommand\arraystretch{1.62}
                \hspace{-10pt}
                \begin{tabular}{ l||c |c |c  }
			\thickhline
            \rowcolor{mygray}
			& \makecell{Speech Enhancement \\ PESQ $\uparrow$}  & \makecell{Speech Recognition \\ WER($\%$) $\downarrow$}  & \makecell{Speaker Verification\\ EER($\%$) $\downarrow$}\\
			\hline
			\hline
			Anechoic (Ceiling)                                          & 4.64 & 2.50  & 1.89 \\
			\hline
			Reverberant                                                  & 1.54 & 8.86  & 5.23 \\
			MetricGAN+~\cite{DBLP:conf/interspeech/FuYHPRL021}           & 2.33  & 7.49   & 5.16   \\
            VIDA~\cite{ChenSHG23}                 & 2.37  & 4.44  & 4.58   \\
            \cdashline{1-4}[1pt/1pt]
			\textbf{MVSD}                                                           & \bfseries 2.53   & \bfseries 4.27  & \bfseries 4.46 \\
			\hline
	       \end{tabular}
        }
        \label{tab:comparison} 
        \end{minipage}
        \begin{minipage}[t]{0.42\textwidth}
            \makeatletter\def\@captype{table}\captionsetup{width=.9\linewidth}
            \centering\small
            \caption{User study results. X\%/Y\% means that X\% of participants prefer this method while Y\% prefer MVSD (\S\ref{exp:US}).}\label{tbl:US}
            \resizebox{1.\linewidth}{!}{
                \setlength\tabcolsep{2.5pt}
                \renewcommand\arraystretch{1.31}
                \begin{tabular}{l ||c |c }
                \thickhline
                \rowcolor{mygray}
                         & SoundSpaces & AVSpeech \\
                \hline
                \hline
                Input Speech            &  39.3\% / \bfseries 60.7\%  & 38.2\% / \bfseries 61.8\%  \\
                Image2Reverb~\cite{singh2021image2reverb}             &  20.8\% / \bfseries 79.2\%  &  - / - \\
                AV U-Net~\cite{DBLP:conf/cvpr/GaoG19}           &  23.4\% / \bfseries 76.6\%  &  21.9\% / \bfseries 78.1\% \\
                AViTAR~\cite{chen22vam}  &  34.7\% / \bfseries 65.3\%  &  44.1\% / \bfseries 55.9\% \\
                \hline
        		\end{tabular}
            }
        \end{minipage}
\vspace{-3pt}
\end{figure*}
\subsection{Performance on Derverberation} \label{PD}
\begin{figure*}
  \centering
  \includegraphics[width=1 \linewidth, trim=20bp 22bp 20bp 22bp, clip]{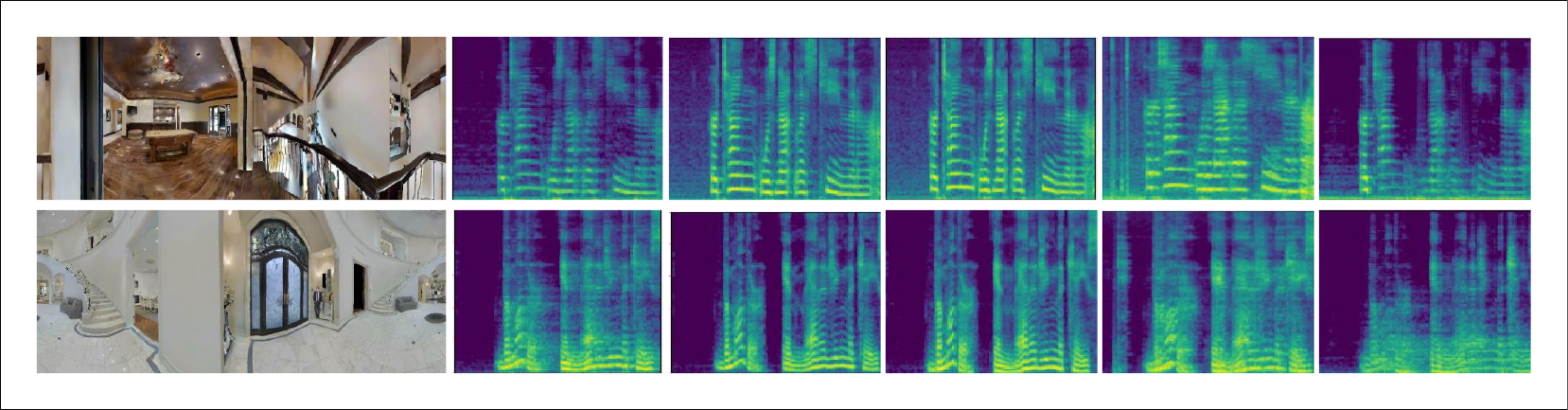}
  \put(-318,78){\scalebox{.80}{\texttt{Visual scene}}}
  \put(-256,78){\scalebox{.75}{\texttt{Reverberant audio}}}
  \put(-180,78){\scalebox{.75}{\texttt{GT}}}
  \put(-143,78){\scalebox{.75}{\texttt{MVSD\,(Ours)}}}
  \put(-87,78){\scalebox{.75}{\texttt{VIDA$\!\!\!\!$\cite{ChenSHG23}}}}
  \put(-48,78){\scalebox{.75}{\texttt{MetricGAN+$\!\!\!\!$\cite{DBLP:conf/interspeech/FuYHPRL021}}}}
\caption{Visualization of the dereverberation task on the SoundSpaces-Speech~\cite{chen22vam}~(\S\ref{PD}). }
\label{fig:VisDereverb}
\vspace{-13pt}
\end{figure*}

Table~\ref{tab:comparison} presents the dereverberation performance of MVSD on SoundSpaces-Speech\!~\cite{chen22vam} dataset.
We observe that MVSD also demonstrates superior performance across all three metrics in the dereverberation task.
Particularly in terms of WER, MVSD exhibits a remarkable error reduction of $0.17$ compared to VIDA, achieving a value of $4.27\%$.
This highlights the robust dereverberation capability of MVSD.
Additionally, MVSD achieves an EER of $4.46\%$, demonstrating its ability to mitigate reverberation while preserving the timbre information.
Fig.~\ref{fig:VisDereverb} depicts the spectrograms for the dereverberation task on SoundSpaces-Speech, with AVSpeech dataset omitted due to the absence of groundtruth anechoic audio.
Spectrogram analysis reveals that MVSD achieves superior clarity and noise reduction in dereverberation, with distinct peaks and fewer artifacts.
The improvements observed in both tasks signify that MVSD can leverage the reciprocity to enhance the learning capability.
\subsection{User Study} \label{exp:US}
The human ear is the most accurate tool for evaluating acoustic experiences.
Therefore, we conduct a user study as a complement to quantitative indicators.
We invited $15$ volunteers to participate in the evaluation.
Following the configuration in~\cite{chen22vam}, we show participants some images of the target environment, real audio clips, and samples generated by all test methods.
Participants is asked to choose the audio sample that exhibit the highest consistency with the groundtruth reverb style and $30$ samples are selected for each dataset.
Table~\ref{tbl:US}~reports the final preference scores.
As expected, MVSD consistently exceeds the results of other methods, achieving a high preference ratio against AViTAR~\cite{chen22vam}~($65.3\%$ in SoundSpaces and $55.9\%$ in AVSpeech).
A certain percentage ($39.3\%$ and $38.2\%$) of subjects prefer `clean' audio devoid of reverberation, which can be seen from the first row of Table~\ref{tbl:US}.
This tendency can be attributed to the general preference for `clean' sounding audio among those participants without professional acoustical knowledge, which is also reported in~\cite{chen22vam}.
\subsection{Ablation Study} \label{AS}
To assess the effectiveness of MVSD's key components, we conduct diagnostic studies and report VAM results on SoundSpaces-Speech dataset\!~\cite{chen22vam} and using WER and PESQ metrics for dereverberation.

\noindent\textbf{Mutual$_{\!}$ Learning.$_{\!}$}
We$_{\!}$ conduct$_{\!}$ three$_{\!}$ diagnostic$_{\!}$ experiments:$_{\!}$
i) VSD -- training two tasks separately with the structure akin to mutual learning;
ii) MVSD~\textit{w/o} unpaired data -- training exclusively with labeled data; and
iii) MVSD -- augmenting the second experiment with unpaired data.
Table~\ref{tab:ML}~reveals that our baseline model VSD can achieve performance matching SOTA on metrics STFT~$0.657$ and MOSE~$0.159$.
The introduction of MVSD results in a slight edge over SOTA, and incorporating unpaired data notably surpasses SOTA in both tasks~(achieved $0.508$~STFT, $0.030$~RTE, $0.142$~MOSE in VAM, and $4.27\%$~WER, $2.53$~PESQ in dereverberation, respectively).
These findings highlight that the synergy between dual tasks can enhance learning capabilities and efficiently absorb unpaired data, showcasing the benefit of a wealth of natural data.

\noindent\textbf{Model Design.} 
To validate the superiority of diffusion model, we conduct comparative experiments with two different generator architectures:
(1) a conditional generative network based on GAN, influenced by our single-task model, and
(2) The controllable Unet in MVSD, designed to showcase the diffusion process.
Table~\ref{tab:DM} shows the controllable Unet outperforms the GAN-based model significantly in all evaluated metrics,~STFT reduced by $0.078$ to $0.753$. WER decreased by $1.57\%$ to $6.74\%$.
The diffusion process significantly further enhances the performance to achieve $0.657$ STFT and $4.27\%$ WER.
Compared to GANs, diffusion models can excel in stability and sample quality, enabling a more controllable and precise generation process.
\begin{table*}[t]
\caption{A series of ablation studies on SoundSpaces-Speech dataset\!~\cite{chen22vam}~(\S\ref{AS}).}
    \label{tbl:AS}
    \begin{minipage}{\textwidth}
        \begin{subtable}[t]{0.5\textwidth}
            \makeatletter\def\@captype{table}\captionsetup{width=.9\linewidth}
            \centering\small
            \resizebox{1.\linewidth}{!}{
                \setlength\tabcolsep{1pt}
                \renewcommand\arraystretch{1.22}
                \begin{tabular}{ l||c |c | c ||c|c }
			\thickhline
            \rowcolor{mygray}
            Method & \multicolumn{3}{c||}{\textit{VAM}}                                                                                                       & \multicolumn{2}{c}{\textit{Derverberation}}  \\
            \rowcolor{mygray}
                               & STFT~$\downarrow$ & RTE(s)~$\downarrow$ & MOSE~$\downarrow$ & WER~$\downarrow$ & PESQ~$\uparrow$\\
			\hline
			\hline
            VSD       & 0.657 & 0.037  & 0.159  &  4.39 &   2.41  \\
			MVSD~\textit{w/o} unpaired data  & 0.573 & 0.033 & 0.148   & 4.32  &  2.47  \\
			\textbf{MVSD}          & \bfseries 0.508 & \bfseries 0.030 & \bfseries 0.142  & \bfseries 4.27 & \bfseries 2.53  \\
			\hline
			
	       \end{tabular}
        }
        \caption{mutual learning}
        \label{tab:ML}
        \end{subtable}
        \begin{subtable}[t]{0.5\textwidth}
            \makeatletter\def\@captype{table}\captionsetup{width=.9\linewidth}
            \centering\small
            \resizebox{1.\linewidth}{!}{
                \setlength\tabcolsep{2pt}
                \renewcommand\arraystretch{1.2}
                \begin{tabular}{l ||c |c |c ||c|c }
                \thickhline
                \rowcolor{mygray}
                Method & \multicolumn{3}{c||}{\textit{VAM}}                                                                                                       & \multicolumn{2}{c}{\textit{Derverberation}}  \\
                \rowcolor{mygray}
                  & STFT $\downarrow$ & RTE(s) $\downarrow$ & MOSE $\downarrow$ & WER  $\downarrow$ & PESQ  $\uparrow$ \\
                \hline
                \hline
                CNN-GAN               &  0.831           & 0.076           & 0.237           &  8.31   &    1.93   \\
                Unet~\textit{w/o} diffusion        &  0.753           & 0.067           & 0.194         &   6.74    &  2.19          \\
                \textbf{Diffusion}      & \bfseries 0.657  & \bfseries 0.037   & \bfseries 0.159  & \bfseries 4.27 & \bfseries 2.53 \\
                \hline
        		\end{tabular}
            }
            \caption{diffusion model}
            \label{tab:DM}
        \end{subtable}
    \end{minipage}
    \begin{minipage}{\textwidth}
        \begin{subtable}[t]{0.5\textwidth}
            \makeatletter\def\@captype{table}\captionsetup{width=.9\linewidth}
            \centering 
            \small
            \resizebox{1.\linewidth}{!}{
                \setlength\tabcolsep{2.0pt}
                \renewcommand\arraystretch{1.03}
                \begin{tabular}{l ||c |c |c ||c|c||c}
                \thickhline
                \rowcolor{mygray}
                & \multicolumn{3}{c||}{\textit{VAM}}                                                                                                       & \multicolumn{2}{c||}{\textit{Derverberation}} & Timeliness\\
                \rowcolor{mygray}
                Steps & STFT  $\downarrow$ & RTE(s)  $\downarrow$ & MOSE  $\downarrow$  & WER $\downarrow$ & PESQ  $\uparrow$ & RTF $\uparrow$\\
                \hline
                \hline
                150             &  1.452    & 0.242   & 0.376               &  8.39  &  1.48   & \textbf{0.253} \\
                250             &  0.508    & 0.030   & 0.142               &  4.27  &  2.53  &  0.426 \\
                350             &  0.493    & 0.035   & \textbf{0.139}     &  \textbf{4.26}  &  2.47  &  0.619 \\
                500             &  0.492    & \textbf{0.029}   & 0.144      & 4.28   & \textbf{2.55}  & 0.898  \\
                1000            &  \textbf{0.487}    & 0.033   &  0.141    &  4.35 &  2.49  &  1.809 \\
                \hline
        		\end{tabular}
        }
        \caption{denoising steps}
        \label{tab:DS}
        \end{subtable}
        \begin{subtable}[t]{0.5\textwidth}
            \makeatletter\def\@captype{table}\captionsetup{width=.94\linewidth}
            \centering\small
            \resizebox{1.\linewidth}{!}{
                \setlength\tabcolsep{2pt}
                \renewcommand\arraystretch{1.24}
                \begin{tabular}{ l||c|c |c||c|c}
			\thickhline
            \rowcolor{mygray}
                & \multicolumn{3}{c||}{\textit{VAM}}                                                                                                       & \multicolumn{2}{c}{\textit{Derverberation}}  \\
            \rowcolor{mygray}
			Num & STFT  $\downarrow$ & RTE(s)  $\downarrow$ & MOSE  $\downarrow$ & WER $\downarrow$ & PESQ  $\uparrow$ \\
			\hline
			\hline
			$0~k$          &  0.573   &  0.033   &  0.148                           & 4.32 &  2.47 \\
			$1~k$          &  0.547(\scalebox{.70}{+4.5\%})  &  0.033(\scalebox{.70}{+0.0\%})  &  0.147(\scalebox{.70}{+0.7\%})    & 4.28 &  2.50     \\
			$3~k$          &  0.521(\scalebox{.70}{+9.1\%})  &  0.031(\scalebox{.70}{+6.1\%})  &  0.143(\scalebox{.70}{+3.4\%})  & 4.29 &  2.52    \\
            \boldsymbol{$5~k$}          & \bfseries 0.508(\scalebox{.70}{+11.3\%})  & \bfseries 0.030(\scalebox{.70}{+9.1\%})  & \bfseries 0.142(\scalebox{.70}{+4.1\%}) & \bfseries 4.27 & \bfseries 2.53    \\
			\hline
	       \end{tabular}
            }
            \caption{unpaired data size}
            \label{tab:unpair}
        \end{subtable}
    \end{minipage}
\vspace{-18pt}
\end{table*}

\noindent\textbf{Denoising Steps.} 
We conduct experiments varying the number of denoising steps, as detailed in Table~\ref{tab:DS} and apply the Real-Time Factor~(RTF) to measure the speed of audio generation relative to the actual duration of the audio.
Results indicate suboptimal generation for steps under $250$. At $250$ steps, MVSD matches SOTA performance.
However, more steps require longer training time but yield minimal improvement, increasing the runtime by at least $30\%$.
Consequently, we establish $250$ as the optimal number of diffusion steps.

\noindent\textbf{Unpaired Data Size.} 
We diagnose the impact  of unlabeled~data by investigating the correlation between the quantity of unpaired data and performance.
As shown in Table~\ref{tab:unpair}, increasing the amount of unpaired data consistently boosts the performance.
Incorporating unpaired data equivalent to $17.3\%$ of the supervised samples, STFT-distance shows a notable improvement of $11.3\%$.
Similar conclusions can also be observed in the dereverberation results.
More unpaired data enables the model to learn from a broader data distribution, improving its predictive accuracy and stability.
\section{Conclusion}
\label{sec:con}
In this paper, we introduce MVSD, a mutual learning framework based on visual scene-driven diffusion model, designed for VAM and dereverberation tasks.
In early exploration, we combine diffusion model with mutual learning, a strategy that leverages the complementary aspects between tasks to improve both the performance and the generalization capabilities. 
Consequently, MVSD achieves SOTA performance in the both tasks.
We empirically demonstrate that by utilizing a symmetric diffusion model architecture, MVSD can effectively extract and utilize cross-task knowledge across both tasks.
Furthermore, by integrating an additional $17.3\%$ of unpaired data into the training set, we have observed a $9.1\%$ relative improvement in RTE for VAM.
This strategy allows MVSD to access easily acquired unpaired data, thereby reducing the reliance on annotation.
We anticipate our research will enhance the utilization of unidirectional data.
\clearpage
\bibliographystyle{splncs04}
\bibliography{main}

\clearpage
\appendix
\section*{\centering \textbf{Supplementary Materials}}
\setcounter{table}{0}
\setcounter{figure}{0}
\setcounter{footnote}{0}
\renewcommand{\thetable}{A\arabic{table}}
\renewcommand{\thefigure}{A\arabic{figure}}
\renewcommand{\thepage}{A\arabic{page}}
In the appendix, we provide the following content for a more comprehensive understanding of our method:
\begin{itemize}[leftmargin=*]
	\setlength{\itemsep}{0pt}
	\setlength{\parsep}{-2pt}
	\setlength{\parskip}{-0pt}
	\setlength{\leftmargin}{-10pt}
  \item \S~\ref{sec:AD}: \textbf{Architecture Details}. We provide details of the MVSD network architecture, including layer composition, connectivity patterns, \etc.
  \item \S~\ref{sec:LSI}: \textbf{Social Impacts and Limitations}.
  \item \S~\ref{sec:QV}: \textbf{Qualitative Visualization} of MVSD and several competitors.
\end{itemize}
\begin{figure*}[b]
  \centering
      \includegraphics[width=1 \linewidth, trim=15bp 10bp 15bp 10bp, clip]{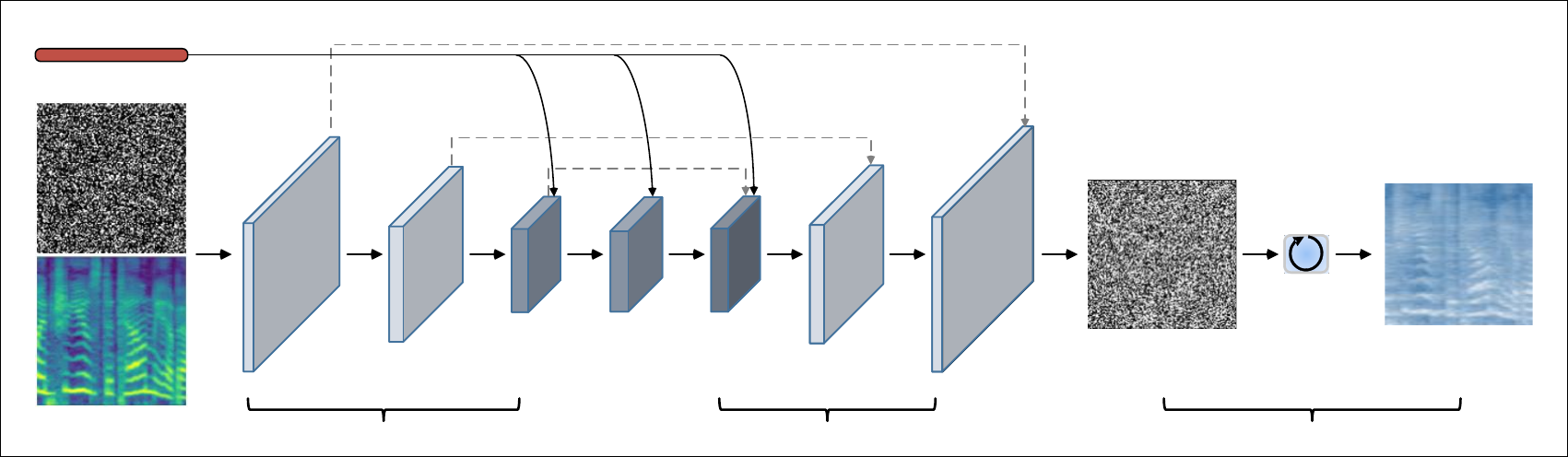}
      \put(-343,80){\scalebox{.65}{$F_v:1\times256$}}
      \put(-343,0){\scalebox{.65}{$z_T:128^2\times1$}}
      \put(-309,10){\scalebox{.65}{$32^2\times128$}}
      \put(-273,17){\scalebox{.65}{$8^2\times256$}}
      \put(-247,23){\scalebox{.65}{$4^2\times512$}}
      \put(-221,23){\scalebox{.65}{$4^2\times512$}}
      \put(-195,23){\scalebox{.65}{$4^2\times512$}}
      \put(-176,16){\scalebox{.65}{$8^2\times256$}}
      \put(-149,10){\scalebox{.65}{$32^2\times128$}}
      \put(-105,18){\scalebox{.65}{$z_{T-1}:128^2\times1$}}
      \put(-66,32){\scalebox{.65}{$\times(T-1)$}}
      \put(-22,21){\scalebox{.65}{$z_0$}}
      \put(-274,-2){\scalebox{.75}{Encoder}}
      \put(-173,-2){\scalebox{.75}{Decoder}}
      \put(-68,-2){\scalebox{.75}{Denoising step}}
      \put(-41,86){\scalebox{.60}{cyclic iteration}}
      \put(-41,79){\scalebox{.60}{skip connection}}
      
  \vspace{-7pt}
  \caption{Overview of the controllable Unet: MVSD utilizes a frozen visual scene encoder to encode an input RGB image into a visual feature $F_v$ which is then mapped to a $128\times128$ spectrogram by Controllable Unet, facilitating auditory style transformations~(\S\ref{sec:AD}).}
  \vspace{-13pt}
  \label{fig:unet}
\end{figure*}
\section{Architecture Details}
\label{sec:AD}
Our neural network draws inspiration from the Unet structure of Imagen~\cite{DBLP:conf/nips/SahariaCSLWDGLA22}. 
Taking VAM as an example, in each step of diffusion, the controllable Unet learns to perform cross-modal generation using noisy input, clean spectrograms, and embeddings of the visual environment.
As shown in~Fig.~\ref{fig:unet}, we divide controllable Unet into encoder and decoder with symmetric structure and both of them consist of $3$ attention blocks.
Skip connections~\cite{he2016deep} are employed to bridge encoder and decoder, recovering spatial information lost in downsampling.
We only apply cross-modal attention~\cite{VaswaniSPUJGKP17} in the third block of the encoder and the first block of the decoder to connect visual cues and spectrograms.
In self-attention block, we utilize the downsampling module~\cite{DBLP:conf/pkdd/SunkaraL22} with a stride of $4$ to rapidly reduce the size of the feature map. 
The feature map undergoes a size transformation in the controllable Unet~(\small{$128^2\rightarrow32^2\rightarrow8^2\rightarrow4^2\rightarrow8^2\rightarrow32^2\rightarrow128^2$}).
The diffusion training process involves the following steps:
starting with a sample from the data distribution, noise is gradually added over a fixed number of timesteps, creating a sequence of increasingly noisy images to reconstruct the original input.
During inference, the goal is to generate samples from the learned distribution by starting with pure noise and sequentially applying the trained UNet model to denoise the image over timesteps.
\section{Social Impacts and Limitations}
\label{sec:LSI}
MVSD can enrich VR and AR auditory experiences with more realistic acoustics that complement the protagonist's surroundings.
VAM can enhance personalized advertising, assistive technologies, and speech synthesis and recognition.
Furthermore, dereverberation task can boost speech audibility across various environments, reducing reverberation for clearer communication in teleconferencing, broadcasting, and public spaces.
Nevertheless, there are potential risks concerning privacy, possible misuse, and ethics, notably in diverse societal backgrounds.

While MVSD presents a promising potential, there is still scope for further exploration and investigation.
Diffusion models require more time for training and sampling than GANs, posing significant challenges for real-time applications such as meetings and sound rendering, \etc.
Future efforts will focus on  how to integrate methods like~\cite{SongME21} and~\cite{abs-2310-04378} to reduce the number of parameters and speed up diffusion models.

\section{Qualitative Visualization} \label{sec:QV}
\begin{figure*}[h]
\vspace{10pt}
  \centering
  \begin{minipage}{1.\textwidth}
  \centering
    \includegraphics[width=1.\linewidth, trim=20bp 20bp 20bp 20bp, clip]{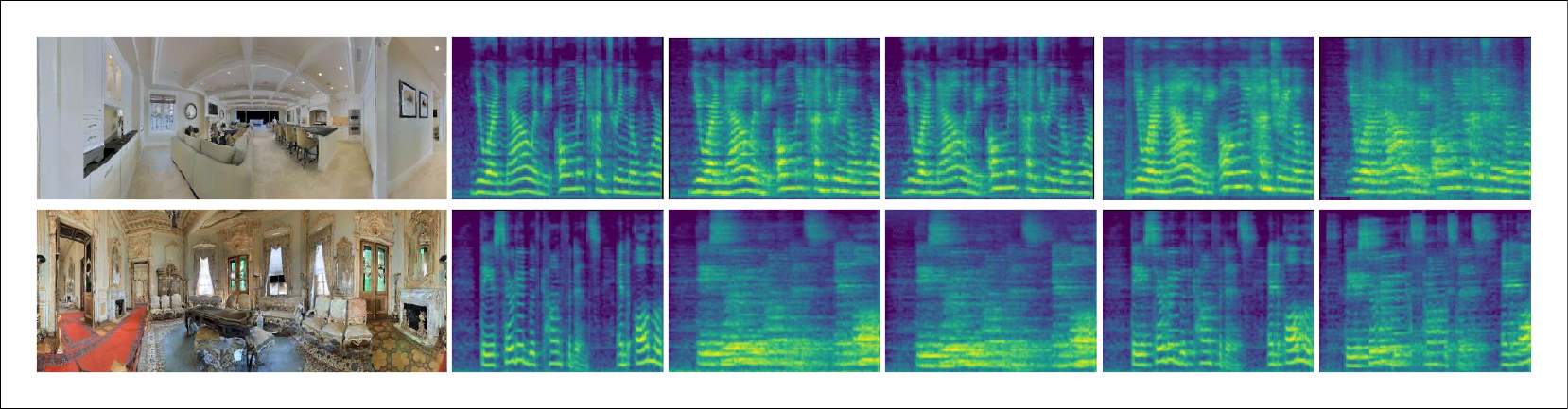}
  \put(-318,80){\scalebox{.80}{\texttt{Visual scene}}}
  \put(-251,80){\scalebox{.75}{\texttt{Anechoic audio}}}
  \put(-180,80){\scalebox{.75}{\texttt{GT}}}  
  \put(-143,80){\scalebox{.75}{\texttt{MVSD\,(Ours)}}}
  \put(-91,80){\scalebox{.75}{\texttt{AViTAR$\!\!\!\!$\cite{chen22vam}}}}
  \put(-51,80){\scalebox{.75}{\texttt{Image2Reverb$\!\!\!\!$\cite{singh2021image2reverb}}}}
    \caption{Visualization results on SoundSpaces-Speech dataset in VAM task~(\S\ref{sec:QV}).}
    \label{fig:ss_reverb}
\end{minipage}
\begin{minipage}{1.\textwidth}
 \vspace{30pt}
  \centering
    \includegraphics[width=1 \linewidth, trim=20bp 20bp 20bp 20bp, clip]{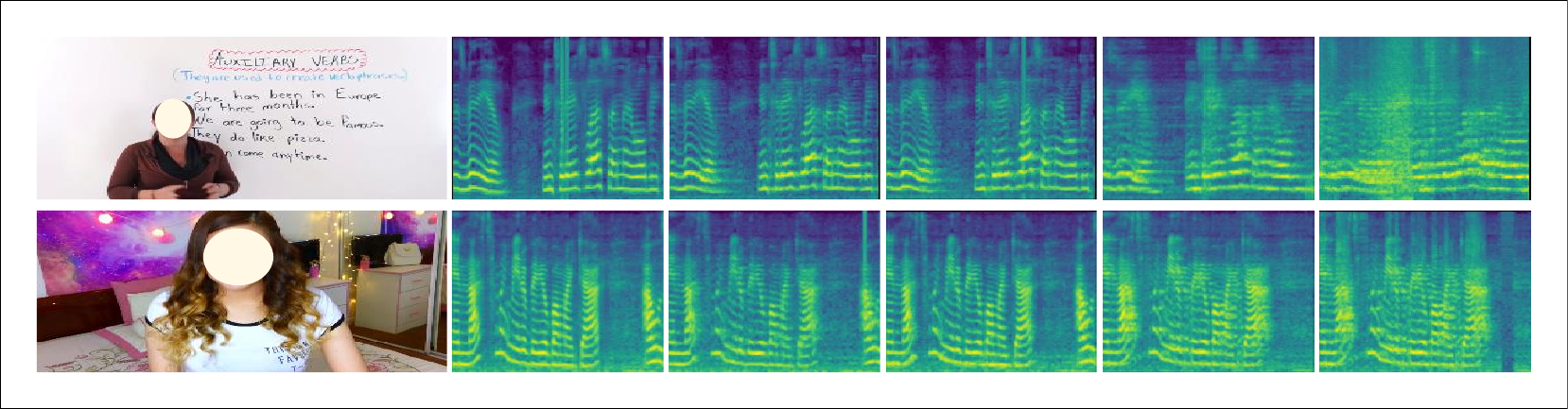}
\put(-318,80){\scalebox{.80}{\texttt{Visual scene}}}
  \put(-251,80){\scalebox{.75}{\texttt{Anechoic audio}}}
  \put(-180,80){\scalebox{.75}{\texttt{GT}}} 
  \put(-143,80){\scalebox{.75}{\texttt{MVSD\,(Ours)}}}
  \put(-91,80){\scalebox{.75}{\texttt{AViTAR$\!\!\!\!$\cite{chen22vam}}}}
  \put(-51,80){\scalebox{.75}{\texttt{Image2Reverb$\!\!\!\!$\cite{singh2021image2reverb}}}}
    \caption{Visualization results on AVSpeech dataset in VAM task~(\S\ref{sec:QV}).}
    \label{fig:avs_reverb}
\end{minipage}
\end{figure*}

\begin{figure*}[t]
  \centering
\begin{minipage}{1.\textwidth}
  \centering
    \includegraphics[width=1 \linewidth, trim=20bp 20bp 20bp 20bp, clip]{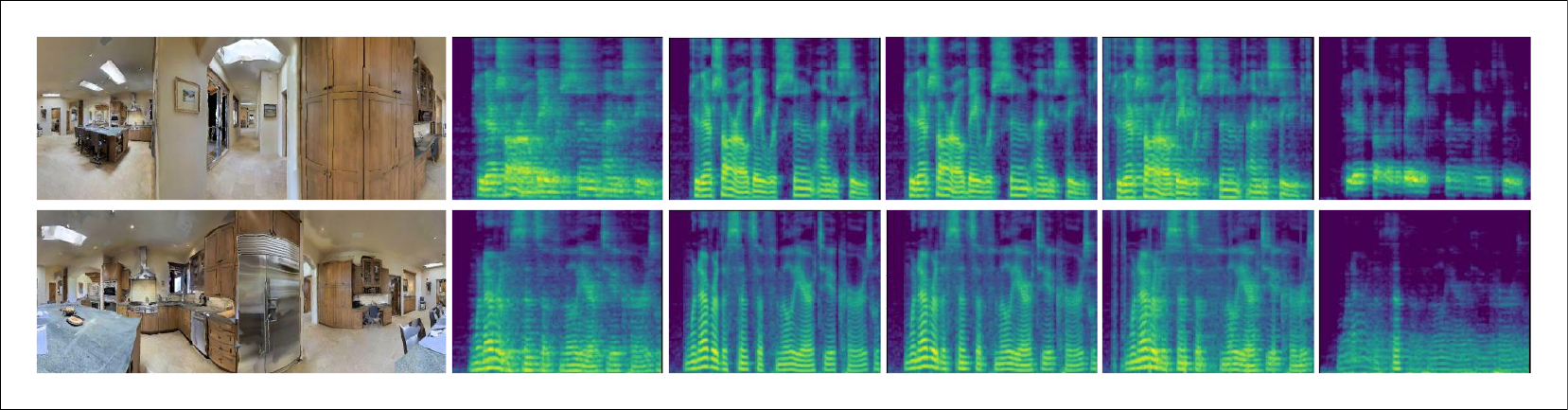}
    \put(-323,80){\scalebox{.80}{\texttt{Visual scene}}}
  \put(-256,80){\scalebox{.75}{\texttt{Reverberant audio}}}
  \put(-180,80){\scalebox{.75}{\texttt{GT}}}
  \put(-143,80){\scalebox{.75}{\texttt{MVSD\,(Ours)}}}
  \put(-87,80){\scalebox{.75}{\texttt{VIDA$\!\!\!\!$\cite{ChenSHG23}}}}
  \put(-46,80){\scalebox{.75}{\texttt{MetricGAN+$\!\!\!\!$\cite{DBLP:conf/interspeech/FuYHPRL021}}}}
    \caption{Visualization on SoundSpaces-Speech dataset in dereverberation task~(\S\ref{sec:QV}).}
    \label{fig:ss_dereverb}
\end{minipage}
\begin{minipage}{1.\textwidth}
\vspace{30pt}
    \centering
      \includegraphics[width=1\linewidth, height=0.23\linewidth,trim=20bp 20bp 20bp 20bp, clip]{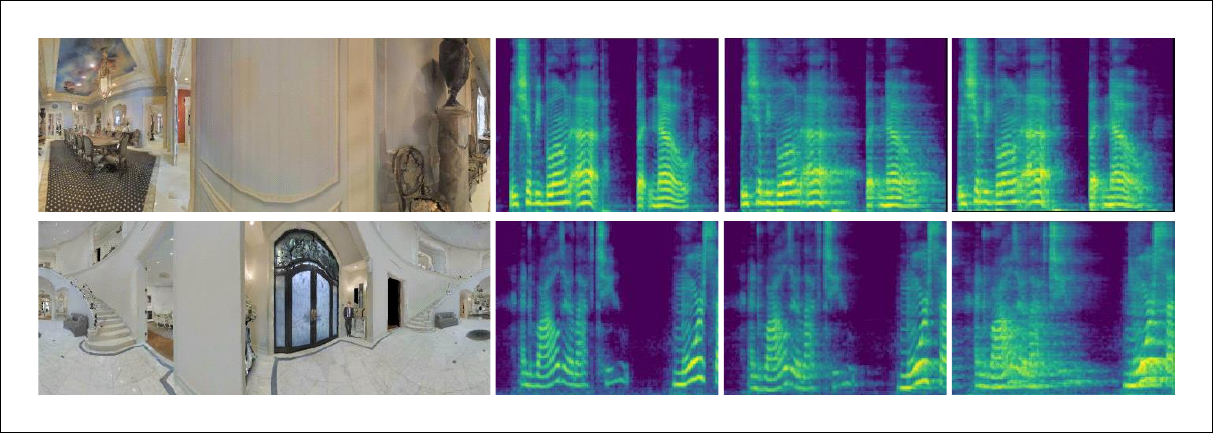}
      \put(-293,83){\scalebox{.70}{\texttt{Visual scene}}}
      \put(-195,83){\scalebox{.70}{\texttt{Anechoic audio}}}
      \put(-105,83){\scalebox{.70}{\texttt{GT}}}
      \put(-41,83){\scalebox{.70}{\texttt{MVSD}}}
        \caption{Some failed samples from SoundSpaces-Speech dataset in VAM task~(\S\ref{sec:QV}).}
      \label{fig:failed}
\end{minipage}
\end{figure*}

\vspace{-3pt}
This section showcases visualizations of qualitative results for our MVSD and competing methods.
Among them, Fig.~\ref{fig:ss_reverb}~and Fig.~\ref{fig:avs_reverb}~depict the qualitative results of the VAM task on different datasets.
Fig.~\ref{fig:ss_dereverb}~showcases the visualization of the generated results on SoundSpaces-Speech dataset in the dereverberation task.
Fig.~\ref{fig:failed}~illustrates some instances of failure cases observed on SoundSpaces-Speech dataset~\cite{chen22vam}.
\end{document}